\documentclass[12pt]{article} 

\renewcommand{\thefootnote}{\fnsymbol{footnote}}

\makeatletter
\newsavebox{\@brx}
\newcommand{\llangle}[1][]{\savebox{\@brx}{\(\m@th{#1\langle}\)}%
  \mathopen{\copy\@brx\mkern2mu\kern-0.9\wd\@brx\usebox{\@brx}}}
\newcommand{\rrangle}[1][]{\savebox{\@brx}{\(\m@th{#1\rangle}\)}%
  \mathclose{\copy\@brx\mkern2mu\kern-0.9\wd\@brx\usebox{\@brx}}}
\makeatother

\usepackage{amsbsy,amssymb,latexsym,amsfonts,amsmath}
\usepackage{mathrsfs}
\usepackage{graphicx}
\usepackage{bm}
\usepackage{color}

\newcommand{\bel}[1]{\begin{equation}\label{#1}}                     
\newcommand{\bal}[1]{\begin{eqnarray}\label{#1}}                     
\newcommand{\be}{\begin{equation}}
\newcommand{\ee}{\end{equation}}

\newcommand{\nn}{\nonumber }

\begin{document}
%
%
\begin{titlepage}
\begin{flushright}
\normalsize
~~~~
NITEP 275
April 16 2026
\end{flushright}

\vspace{15pt}

\begin{center}
	
{\LARGE Fermionic modes of D-instanton wormholes \\ from broken local supersymmetry} \\
\vspace{10pt}
\end{center}

\vspace{23pt}

\begin{center}
{H. Itoyama$^{a}$\footnote{e-mail: itoyama@omu.ac.jp},
Hikaru Kawai$^{a}$\footnote{e-mail: hikarukawai@phys.ntu.edu.tw}, 
and Shoichi Kawamoto$^{b,a}$ \footnote{e-mail: kawamoto\_s@obirin.ac.jp}  }\\

\vspace{18pt}


$^a$ \textit{Nambu Yoichiro Institute of Theoretical and Experimental Physics (NITEP), \\
Osaka Metropolitan University, 
3-3-138, Sugimoto, Sumiyoshi-ku, Osaka, 558-8585, Japan} \\



$^b$ \it  College of Arts and Sciences, J.~F.~Oberlin University, Tokyo, Japan\\

\end{center}

\vspace{20pt}

\begin{center}
Abstract\\
\end{center}

 In low-energy supergravity treatment of type IIB superstring on general D-instanton wormhole profiles in the bulk, we obtain non-vanishing scalar two-point functions in addition to the vanishing $\llangle \tau^* \tau^* \rrangle$ that corresponds to the BPS amplitude detected by two D-instantons at their respective boundaries. This is exploited to show that the modes of broken local supersymmetry in the bulk deliver the fermionic (diagonal) modes on the boundaries through the deformation by the form of current-current two point functions  propagating on the tree level cylinder geometry. Our treatment is generalizable to multi D-instanton cases and general Euclidean branes.
  

\vfill

\end{titlepage}

\renewcommand{\thefootnote}{\arabic{footnote}}
\setcounter{footnote}{0}

\paragraph{Introduction}

Interplay between the closed/open string pictures has made several hallmarks in string theory ever since its birth \cite{GSW}. In the last three decades, the low-energy supergravity description and that of supersymmetric-Yang Mills have been effective in revealing nonlinear superstring dynamics from their respective closed and open string pictures.
   In this letter, we take the closed string description to investigate the role played by the fermionic modes associated with broken local supersymmetry in the bulk on the dilaton-axion profiles of D-instantons in the type IIB  supergravity. This is found to bring the boundaries those fermionic modes that can be identified with the diagonal Grassman elements of zero dimensional super Yang-Mills theory, namely, the matrix models of this type \cite{Banks:1996vh, Ishibashi:1996xs, Dijkgraaf:1997vv, Itoyama:1997gm, Chepelev:1997av}, which are elusive and yet change the dynamics on the boundaries drastically \cite{Green:1996xf, Suyama:1997ig, Aoki:1998vn}.\footnote{%
Other aspects of D-instantons include \cite{Witten:1995im, Green:1997tv, Green:1997tn}.}
Searching for the counterpart of such degrees of freedom coming from the bulk, we deform the action of IIB supergravity at a linearized level by these modes of the broken local supersymmetry
to obtain $S_\text{eff}$ in the form of the two-point function of supercurrents propagating on cylinder. 

In order to set a starting point of our consideration, let us first recall how the vanishing open string annulus superstring amplitudes and the open-closed string duality have led us to the BPS cancellation of the cylinder amplitude with two of D-instantons being the initial/final boundary states at $X_{1}$ and at $X_{2}$:
\begin{align}
\label{eq:BPS_cancellation1}
  0=& \int dt \langle B; X_2 | e^{-Ht} |B; X_1 \rangle
=\int dt  \int d^{10}x \int d^{10}x' \langle B;X_2 | x \rangle \langle x | e^{-Ht} | x'\rangle \langle x' | B; X_1 \rangle \,.
\end{align}
At large Euclidean distance $r=|X_{1} - X_{2}|$, this reads the BPS cancellation between the propagators ($\sim r^{-8}$) of bosonic massless particles \cite{Polchinski:1995mt}. 
As D-instantons can couple only to functions (zero-forms), their profiles are shown at dilatons $\phi$ in NS$\otimes$NS and the axion $a$ in R$\otimes$R, which are organized as $\tau(a,\phi)=a+ie^{-\phi}$ and $\tau^*(a,\phi)=a-ie^{-\phi}$.
A noteworthy property is that, after Euclideanization $a=-i\alpha$, the axion has a wrong sign in its kinetic term. This is accounted for by the sensitivity of p forms in general to
the signature factor of space time.

In the low energy supergravity approximation, we regard $\langle x | B; X_{1,2} \rangle$ to consist of $\mathbf{8_{NS^2}\oplus 8_{R^2}}$ real bosonic delta function sources that couple respectively to the dilaton and the axion, and \eqref{eq:BPS_cancellation1} reduces to
\begin{align}
=&
\iint d^{10}xd^{10}x' \big(j_{\mathbf{8_{NS^2}}}(x|X_2), j_{\mathbf{8_{R^2}}}(x|X_2) \big)
        \begin{pmatrix}
          1 & 0 \\
          0 & -1
        \end{pmatrix}
\hat{D}_x \delta^{(10)}(x-x') 
        \begin{pmatrix}
          j_{\mathbf{8_{NS^2}}}(x'|X_1)  \\ j_{\mathbf{8_{R^2}}}(x'|X_1) 
        \end{pmatrix}
\nn\\ \equiv &
\iint d^{10}xd^{10}x'
\bm{j}_{\mathbf{8_{NS^2}\oplus 8_{R^2}}} (x|X_2) \cdot \bm{D}_{x, \text{dilaton-axion}}
\delta^{(10)}(x-x') 
\bm{j}_{\mathbf{8_{NS^2}\oplus 8_{R^2}}}(x'|X_1) 
\nn\\ \equiv &
\bm{j} \cdot \hat{\bm{D}} \bm{j} 
\equiv {\cal C}_{\mathbf{8_{NS^2}\oplus 8_{R^2}}}(X_2, X_1)  \,,
\label{eq:BPS_cancellation2}
\end{align}
Here, $\hat{D}_x$ is the position space scalar propagator in ten Euclidean dimensions,
\begin{align}
  \bm{j}_{\mathbf{8_{NS^2}\oplus 8_{R^2}}}(x|X) =
  \begin{pmatrix}
    \langle x | B ; X, \mathbf{8_{NS^2}} \rangle \\
    \langle x | B ; X, \mathbf{8_{R^2}} \rangle
  \end{pmatrix} \,,
\end{align}
a pair of delta function sources with reality assumed in the bracket $\langle \phantom{a} , \phantom{a} \rangle$, and
\begin{align}
  \hat{\bm{D}} =
  \begin{pmatrix}
    \hat{D} & 0 \\
    0 & - \hat{D}
  \end{pmatrix} \,.
\end{align}
The equality in strengths of the two sources up to the sign ensures the vanishing amplitude. Alternatively, one can introduce a projection matrix
${\cal P}_\pm=\frac{1}{2}\bigl(\begin{smallmatrix}  1 & \pm 1 \\ \pm 1 & 1 \end{smallmatrix}\bigl)$, ${\cal P}_\pm^2={\cal P}_\pm$, and insert them in two places in between the
 propagator and the (initial and final) sources. Either choice of $\pm$ is admissible for now.

In quantum field theory, $e^{{\cal C}_{\mathbf{8_{NS^2}\oplus 8_{R^2}}}(X_2, X_1)}$ rather than ${\cal C}_{\mathbf{8_{NS^2}\oplus 8_{R^2}}}(X_2, X_1)$ is generated
and \eqref{eq:BPS_cancellation2} is stated as
\begin{align}
\label{eq:one-point1}
  \frac{\langle 0 | 0 \rangle_{\bm{j}} }{\langle 0 | 0 \rangle_{\bm{j}=\bm{0}}}
= e^{{\cal C}_{\mathbf{8_{NS^2}\oplus 8_{R^2}}}(X_2, X_1)} = 1 \,.
\end{align}
In fact, in Euclidean path integrals, we evaluate the numerator as
\begin{align}
\label{eq:expC_1}
  \langle 0 | 0 \rangle_{\bm{j}} =&
\int [{\cal D} \phi] [{\cal D} \alpha ] e^{-S-{{({\cal P}_\pm \bm{j}) \cdot \binom{\phi}{\alpha}}}} \,,\\
S_0 =& \frac{1}{2}\phi \cdot {{ \hat{M}_0 }} \phi + \frac{1}{2} \alpha \cdot (-) {{ \hat{M}_0 }} \alpha \,,
\qquad
 \langle x | \hat{M}_0 | x' \rangle  = (-\partial^2) \delta (x-x') \,.
\end{align}
Here, $\bm{j}_{\mathbf{8_{NS^2}\oplus 8_{R^2}}}= \sum_i \bm{j}_{\mathbf{8_{NS^2}\oplus 8_{R^2}}}(x|X_i)$.
Completing the square with the identification of $\hat{K}^{-1}=\hat{D}$,
we obtain \eqref{eq:one-point1}.

\paragraph{Main analysis}

Having finished the introduction, let us extend this discussion to include the full nonlinear interactions of the low energy type IIB supergravity on the D-instanton classical
 background \cite{GGP} seen as $\phi$, $\alpha$ profiles in Euclidean signature.
The dilaton-axion part of the Lagrangian in the Lorentzian signature is
\begin{align}
  \mathcal{L}_{\phi,a} =& \frac{1}{2 (\text{Im} \tau(a,\phi))^2} \partial_M \tau(a,\phi) \partial^M \tau(a,\phi)^*
= \frac{1}{2} \left( e^{2\phi} (\partial_M a)^2 + (\partial_M \phi)^2 \right) \,,
\end{align}
where $a$ is the axion in the Lorentzian signature.
The axion in the Euclidean signature is given by $a=-i\alpha$ and
the corresponding Euclidean Lagrangian reads
\begin{align}
\mathcal{L}_{\phi,\alpha} =& \frac{1}{2} \left( -e^{2\phi} (\partial_M \alpha)^2 + (\partial_M \phi)^2 \right) 
= \frac{1}{2} \left(-e^\phi (\partial_M \alpha)+\partial_M \phi \right)
              \left(e^\phi (\partial^M \alpha)+\partial^M \phi \right) \,.
\end{align}
The fermionic part will be presented shortly.
The ansatz that we work with is
\begin{align}
  \label{ansatz}
\partial_M \alpha=& \pm  e^{-\phi} \partial_M \phi \,,
\end{align}
while the equation for $\alpha$ and that for $\phi$ are
\begin{align}
  \label{eq:EoM_alpha1}
  \partial_M \big( e^{2\phi} (\partial^M \alpha ) \big) =& 0 \,,\\
  \label{eq:phi_eom1}
  \partial_M \partial^M \phi + e^{2\phi} (\partial_M \alpha ) (\partial^M \alpha ) =& 0 \,,
\end{align}
respectively.
Eqs. \eqref{ansatz}, \eqref{eq:EoM_alpha1} and \eqref{eq:phi_eom1} lead to \cite{GGP}
\begin{align}
  \label{eq:phi_eq2}
  \partial_M \partial^M \big( e^\phi \big) =& 0 \,.
\end{align}
The solutions are trivially generalizable to $n$ D-instantons and are
given by
\begin{align}
\label{eq:dilaton_profile_multi-D}
  e^\phi =& \sum_{i=1}^n \frac{c_i}{|\bm{x}-\bm{X}_i|^8} \,.
\end{align}
This is the dilaton profile and, at the same time, space dependent string coupling constant
and the axion profile is given by \eqref{eq:dilaton_profile_multi-D} up to the constant. It is important to realize that the reality of the dilaton field requires the residues of the solution \eqref{eq:dilaton_profile_multi-D} be all positive.
We conclude that, when the upper sign is adopted in \eqref{ansatz}, we deal with D-instantons only, while, in the case of the lower sign, we deal with anti D-instantons only.
The solution deserves name of wormholes as one sees the presence of the fixed surface after the conversion into the string metric \cite{GGP}.
From now on, we focus on the $n=2$ case.

In perturbation theory of fluctuations on $\phi_c$ and $\alpha_c$ that satisfy eqs. \eqref{ansatz}--\eqref{eq:phi_eq2},  we set
\begin{align}
  \label{eq:fluc_alpha}
  \alpha = & \alpha_c + e^{-\phi_c} \tilde{\alpha} \,,\\
  \label{eq:fluc_phi}
  \phi = & \phi_c + \tilde{\phi} \,.
\end{align}
After some calculation, we obtain the quadratic Lagrangian for the fluctuating fields
$\tilde{{\phi}}, \tilde{{\alpha}}$ into a form which is most manageable to us:
\begin{align}
  \mathcal{L}_2 =&
\frac{1}{2} \big(\tilde{\phi}, \tilde{\alpha} \big)
\begin{pmatrix}
-\Box -2 A_M^2 & \mp 2A_M \partial^M \pm A_M^2 \\                   
\pm 2A_M \partial^M \pm A_M^2 & \Box
\end{pmatrix}
\begin{pmatrix}
\tilde{\phi} \\ \tilde{\alpha}
\end{pmatrix} 
\equiv
\frac{1}{2} \big(\tilde{\phi}, \tilde{\alpha} \big)
M \begin{pmatrix}
\tilde{\phi} \\ \tilde{\alpha}
\end{pmatrix} \,,
\label{eq:L_2_M}
\end{align}
where
\begin{align}
  A_M \equiv \partial_M \phi_c \,,
\quad
M=
  \begin{pmatrix}
    m_{11} & m_{12} \\
    m_{21} & m_{22} 
  \end{pmatrix}
\end{align}
and
\begin{align}
\label{eq:M_element_cancellation1}
  (m_{11}+m_{22} ) \pm (m_{12}+m_{21}) = 0 \,.
\end{align}
This last property \eqref{eq:M_element_cancellation1} is carried over to that of $M^{-1}$:
it is given by
\begin{align}
  \label{eq:M_inv1}
M^{-1} =& 
\begin{pmatrix}
Y^{-1} & \pm Y^{-1} \big( 2A_M \partial^M - A_M^2 \big)\frac{1}{\Box} \\
\mp  \frac{1}{\Box} \big( 2A_M \partial^M + A_M^2 \big)Y^{-1} & 
\frac{1}{\Box} - \frac{1}{\Box} \big( 2A_M \partial^M + A_M^2 \big) Y^{-1}
\big( 2A_M \partial^M - A_M^2 \big) \frac{1}{\Box} 
\end{pmatrix}
\nn\\ \equiv &
\begin{pmatrix}
\tilde{m}_{11} & \tilde{m}_{12} \\
 \tilde{m}_{21} & \tilde{m}_{22}
\end{pmatrix} \,,
\end{align}
where
\begin{align}
  \label{eq:S_def1}
  Y = -\Box -2 A_M^2 
+ \big( 2A_M \partial^M - A_M^2 \big) \frac{1}{\Box} \big( 2A_M \partial^M + A_M^2 \big)
\end{align}
and we check that
\begin{align}
  \label{eq:M-inv_cancellation1}
  (\tilde{m}_{11} + \tilde{m}_{22})\mp (\tilde{m}_{12} + \tilde{m}_{21}) =0 
\end{align}
is satisfied. (Note the sign flip of \eqref{eq:M-inv_cancellation1} from \eqref{eq:M_element_cancellation1}.)

Let us, for definiteness, take the upper sign in the original ansatz \eqref{ansatz}.
The vanishing BPS amplitude (and the two point function) corresponds
to adopting ${\cal P}_-$ projector both at the initial and final sources while remaining three (two kinds) are non-vanishing despite that they are two point functions in instanton backgrounds alone.
To recapitulate this in terms of $\tilde{{\tau}}(a=-i\alpha, \phi) \approx \tilde{\alpha}+\tilde{\phi}$
 and $\tilde\tau^*(a=-i\alpha, \phi) \approx \tilde{\alpha}-\tilde{\phi}$, (tilde meaning the fluctuation or the subtraction of the classical part), ${\cal L}_2 \approx (\tilde{\tau}^*, \tilde{\tau}) 
 \begin{pmatrix}
\# & \# \\ \# & 0
 \end{pmatrix} \binom{\tilde{\tau}^*}{\tilde\tau}$ by \eqref{eq:M_element_cancellation1}.
In this basis,
 \begin{align}
   M^{-1} \approx  \begin{pmatrix}
0 & \#' \\ \#' & \#'
 \end{pmatrix} \,,
 \end{align}
by \eqref{eq:M-inv_cancellation1}. 
Hence, we have one vanishing two-point function
\begin{align}
  \llangle \tilde{\tau}^*(-i\alpha,\phi)(x=X_1)
\tilde{\tau}^*(-i\alpha,\phi)(x=X_2) \rrangle =0 \,,
\end{align}
while the remaining three (two kinds) are nonvanishing:
\begin{align}
    \llangle \tilde{\tau}(-i\alpha,\phi)(x=X_1)
\tilde{\tau}(-i\alpha,\phi)(x=X_2) \rrangle \neq 0 \,,
\quad
  \llangle \tilde{\tau}^*(-i\alpha,\phi)(x=X_1)
\tilde{\tau} (-i\alpha,\phi)(x=X_2) \rrangle \neq 0 \,.
\end{align}
Note that, in accordance with this, the classical solution satisfies
$ \partial_M \tau_{c}^*=0$ while $ \partial_M \tau_{c} \neq 0$.

Having these results at hand, we now look at fermions: our gamma matrix convention in the almost-minus Lorentzian signature reads
\begin{align}
  \label{eq:Gamma_matrices}
  \Gamma^0= \sigma_2 \otimes \mathbf{1}_{16} \,,
\quad \Gamma^i = i \sigma_1 \otimes \gamma^i \quad (i=1,\cdots,9) \,,
\end{align}
where $\gamma^i$ ($i=1,\cdots, 8$) are SO(8) gamma matrices. We take $\gamma^i$ to be real symmetric $16 \times 16$ matrices and $\gamma^9=\gamma^1 \cdots \gamma^8=\text{diag}(\mathbf{1}_8 , -\mathbf{1}_8)$. This choice leads to $\Gamma^{11} = \text{diag}(\mathbf{1}_{16} , -\mathbf{1}_{16})$.
The dilatino $\lambda_{(D)}$ has the positive chirality while the gravitino $\psi_{(D)M}$ and the supersymmetry transformation $\Theta_{(D)}$ have the opposite chirality. Here, we have denoted by subscript (D) 32-component Dirac spinors.  
The fermionic part of the Lagrangian (Lorentzian signature) at the linearized level taken from \cite{Schwarz:1983qr} (see also \cite{Schwarz:1983wa, Bergshoeff:2005ac}) is
\begin{align}
\mathcal{L}_\text{fermions}=& \mathcal{L}_\lambda + \mathcal{L}_\psi + \mathcal{L}_{\text{int}} \,,\\
\label{eq:L_psi}
  \mathcal L_\psi
&= i\,\bar{{\psi}}_{(D)M}\Gamma^{MNL}D_N^{(1/2)}{\psi}_{(D)L},\\[3pt]
\label{eq:L_lambda}
\mathcal L_\lambda
&= i\,\bar{{\lambda}}_{(D)} \Gamma^M D_M^{(3/2)} {\lambda}_{(D)},\\[3pt]
\mathcal L_{\text{int}}
&= -\frac12\,\bar{{\psi}}_{(D)M}\Gamma^N\Gamma^M {\lambda}^*_{(D)}\,P_N
   -\frac12\,\overline{{\lambda}^*}_{(D)} \Gamma^M\Gamma^N{\psi}_{(D)M}\,P_N^* \,.
\label{eq:L_int}
\end{align}
Here, $D_M^{(q)}=\partial_M-iq\,Q_M$  in the flat spacetime and $Q_M$ acts as a gauge field given by
\begin{align}
\label{eq:Q_def}
  Q_M=& \frac{-1}{4\tau_2} \bigg[\frac{\tau-i}{\tau^*-i} \partial_M \tau^* + \frac{\tau^*+i}{\tau+i} \partial_M \tau \bigg] \,.
\end{align}
Also $\tau= \tau_1 + i \tau_2$ and
\begin{align}
  P_M=& -\frac{i}{2\tau_2} Z^* \partial_M \tau \,,
\qquad Z=\frac{\tau+i}{\tau^*-i} \,.
\end{align}

Let us now move on to the transformations, which we have extracted from Ref. \cite{Schwarz:1983qr, Schwarz:1983wa} (converted to $SL(2, \mathbf{R})$ invariant form).
We summarize here the local SUSY transformation laws with ${\Theta}_{(D)}$ being local anticommuting parameter:
\begin{align}
  \delta \tau =& 2i \tau_2 Z  \overline{{\Theta}^*}_{(D)} {\lambda}_{(D)} \,,\\
  \delta {\lambda}_{(D)} =& i \Gamma^M  {\Theta}^*_{(D)} P_M
 \,,\\
\delta {\psi}_{(D)M} =& D_M^{(1/2)} {\Theta}_{(D)} \,.
\end{align}
In \eqref{eq:L_lambda}, we omitted a term irrelevant to our analysis which is 
a linear combination of the field strengths of the NS-NS and R-R two-forms.
Note that $Z$ is a phase factor, $|Z|=1$.

The consequences derived from the classical solutions with regard to broken as well as unbroken supersymmetries are as follows.
As is stated, $\delta \tau_{c}^{*} =0$ and hence $P_{c M}^{*} =0$.
This implies
\begin{align}
\delta \lambda_{(D)}^* = i \Gamma^M \Theta_{(D)} P_{c M}^* =0 \,,
\end{align}
so that $\delta \lambda_{(D)}^*$ is unbroken in perturbation theory on this background.
We see, however, that $\delta \lambda_{(D)}$ is broken. 
In perturbation theory of fluctuation analysis, the second term of \eqref{eq:L_int} is absent:
\begin{align}
\label{eq:L_int_bg}
  \mathcal L_{\text{int}}
&= -\frac12\,\bar{{\psi}}_{(D)M}\Gamma^N\Gamma^M {\lambda}^*_{(D)}\, P_{c N} \,.
\end{align}
By a $U(1)$ gauge rotation of fermions in accordance with \eqref{eq:L_psi}, \eqref{eq:L_lambda}, \eqref{eq:Q_def}, we can absorb the phase of $P_{cN}$ into the redefinition of $Q_M$.
 In order to set up a proper perturbation theory, we require the hermiticity on Eq. \eqref{eq:L_int_bg}:
\begin{align}
\label{eq:reality_cond_Lint}
  \psi_{(D)M}=  \psi_{(D)M}^* \,, \qquad
\lambda_{(D)}=\lambda_{(D)}^* \,,
\end{align}
namely, each being a Majorana-Weyl spinor with its chirality opposite to each other. Eq. \eqref{eq:reality_cond_Lint} coincides with that of type I supergravity.
The upshot is that the relevant fermionic part of Lagrangian to the quadratic order in fluctuations reads
\begin{align}
{\cal L}_\text{fermions}^{fluct.} = &
i\, \bar{\psi}_{\text{(MW)} M} \Gamma^{MNL}\partial_N{\psi}_{\text{(MW)}L}
+ i\, \bar{\lambda}_\text{(MW)} \Gamma^M \partial_M {\lambda}_\text{(MW)} 
\nn\\&
-\frac12\,{\bar{\psi}}_{\text{(MW)}M} \Gamma^N\Gamma^M {\lambda}_\text{(MW)}\, P_{cN} \,,
\label{eq:fermion_fluc_quad}
\end{align}
 where 
${\psi}_{\text{(MW)}M}$ (${\lambda}_\text{(MW)}$) is the real part of $\psi_{(D)M}$ ($\lambda_{(D)}$).
In short, starting from type IIB supergravity, we have obtained \eqref{eq:fermion_fluc_quad} by replacing $P_N$ by $P_{cN}$, which is regarded as the low energy effective action of Nambu-Goldstone fermion associated with local supersymmetry spontaneously broken by D-instantons.
We also have the second pair of Majorana-Weyl fermions which decouples in our analysis and hence do not include them in \eqref{eq:fermion_fluc_quad}.

As is stated in the introduction, the aim of this letter is to see how deformation of the bulk supergravity action delivers the fermionic degrees of freedom on the boundaries, which we will carry out by local supersymmetry. The transformations by the unbroken generators stop as we have seen in perturbation theory and the local supersymmetry ought to be broken to produce such deformation and deliver it to the boundary. Moreover, the annulus-cylinder duality requires that we only pick tree contributions from the bulk and there is no way for the gauge invariant current-current two point functions of unbroken symmetry in general to produce such tree contribution; in the unbroken case (typically in QED), the current begins with quadratic level in fields and no single particle pole contribution at the tree level develops.

At this moment, we can also identify the local Grassman transformation parameters at $X_i$ with the fermionic coordinates denoted by $\Theta_{(D)i} \equiv \Theta_{(D)}(X_{i})$ on general grounds. The argument goes as follows:
 \begin{align}
&   \int d^{10}x \sum_i \delta(x-X_i) \text{(ANY)}
\nn\\ & \rightarrow 
\int d^{10}x \sum_i \delta(x-X_i)\big[1+\Theta_{(D)}(x)Q^*+\Theta_{(D)}^*(x)Q \big]\times \text{(ANY)}
\nn\\=&
\int d^{10}x d\theta d\theta^* \sum_i \delta(x-X_i) 
\delta(\theta-\Theta_{(D)}(X_i)) \delta(\theta^*-\Theta_{(D)}^*(X_i))
\big[1+\theta Q^*(x)+\theta^*Q(x) \big]\times \text{(ANY)} \,,
\label{eq:ANY_eq}
 \end{align}
where we have denoted  by $Q(x)\times$ and $Q^*(x)\times$ supersymmetry transformation law and by (ANY) any local fields including composite ones. We choose it to be the tree action of IIB supergravity below.

With the help of these considerations, we carry out deformation by the following steps.
 First, we introduce localized sources $j_s(x; X_i)$ containing a set of delta functions (which we will not write explicitly) and we regard them as local Grassman and infinitesimal parameters and couple them to fields through the transformation. Second, we deform the action rather than individual fields to ensure gauge invariant procedure. 
This also teaches us to transform so called ``charged fields'' only, leaving aside the gauge fields, namely the gravitinos. (This procedure, well known in the derivation of Ward-Takahashi identity, has been tested by us in the abelian Higgs model to give a proper expression consisting of tree contribution.)
Omitting the subscript (D) from now on, we obtain the effective action for $(X_i, \Theta(X_i))$ by
\begin{align}
Z[\Theta, {\Theta}^*]
=& 
\int 
[\mathcal{D}\lambda] [\mathcal{D}\lambda^*] [\mathcal{D} \tau] [\mathcal{D} \tau^*]
[\mathcal{D} \psi_{M}] [\mathcal{D} \psi_{M}^*]
e^{i S_\text{IIB SUGRA}[\tau^{[1+\Theta]}  , \tau^{*[1+\Theta^*]}  , \lambda^{[1+\Theta^*]}  , \lambda^{*[1+\Theta]}, \psi_{M},  \psi_{M}^* ]}  \nn\\
=& \int 
[\mathcal{D}\lambda'] \cdots [\mathcal{D} \psi_{M}^{*}]
e^{i\big( S^{(2)}_\text{fluct.} [\tau', \tau^{* \prime}, \lambda', \lambda^{* \prime}, \psi_{M} \psi_{M}^*]+ \int D_M \bar\Theta(x) \mathcal{S}^M(x)\big)}
\label{eq:Z_theta1}
\end{align}
where in the second line the quadratic action $S^{(2)}_\text{fluct.}$ has been introduced in our perturbation theory.\footnote{%
In \eqref{eq:ANY_eq} and \eqref{eq:Z_theta1}, we have regarded $\Theta(x)$ to be infinitesimal. Full treatment of the Nambu-Goldstone phenomenon, which is beyond the scope of this letter, requires higher orders in $\Theta(x)$.}
The supercurrent ${\cal S}^M$ is given by
$\mathcal{S}^M(x) = \frac{1}{2} \Gamma^N \Gamma^M {\lambda}^*_\text{(MW)}\, P_{cN}$.
We are interested in evaluating
\begin{align}
e^{iS_\text{eff}[\Theta]} \equiv & \frac{Z[\Theta]}{Z[0]}\bigg|_{\hbar\rightarrow 0}
= \llangle e^{i\int D_M \bar\Theta \mathcal{S}^M}\rrangle_{\hbar \rightarrow 0}
\label{eq:S_eff_X}
\end{align}
where $\hbar \rightarrow 0$ means that we take tree diagrams only.
Let us introduce
\begin{align}
  \llangle X^2 \rrangle_{\hbar \rightarrow 0} =&
\int d^{10}x d^{10}y \, D_M \bar\Theta(x) \llangle {\cal S}^M(x) {\cal S}^N(y) \rrangle_{\hbar\rightarrow 0} D_N \Theta(y) \,,
\end{align}
and the expression \eqref{eq:S_eff_X} exponentiates to give 
\begin{align}
e^{iS_\text{eff}[\Theta]}  = \exp \bigg(-\frac{1}{2}\llangle X^2 \rrangle_{\hbar \rightarrow 0} \bigg) \,.
\end{align}
The reason for this exponentiation goes as follows: expansion of the right hand side of \eqref{eq:S_eff_X} leads to $\llangle (-X^2)^m \rrangle = (-1)^m \llangle X^2 \rrangle \cdots \llangle X^2 \rrangle c_m$ with $c_m=\frac{1}{m!} {}_{2m}C_{2m-2} \cdots {}_2C_2=\frac{(2m)!}{2^m m!}$, where the factor $1/m!$ comes from the fact that $m$ pairs of $\llangle X^2 \rrangle $ are indistinguishable.
Finally, let us evaluate $\llangle X^2 \rrangle_{\hbar \rightarrow 0}$. Since we treat only tree diagrams, we just need $\llangle {\lambda}_\text{(MW)}(x) {\lambda}_\text{(MW)}^T(y) \rrangle$.
From \eqref{eq:fermion_fluc_quad}, it is given by
\begin{align}
   \llangle {\lambda}_\text{(MW)}(x) {\lambda}_\text{(MW)}^T(y) \rrangle
= \bigg[i \Gamma^0\Gamma^M \partial_M 
-\frac{1}{16} P_{cN} \Gamma^0 \Gamma^N \Gamma^M (K_\psi^{(\xi)-1})_{ML} \Gamma^0 \Gamma^L \Gamma^R P_{cR} + \cdots \bigg]^{-1}(x,y) \,,
\end{align}
where $(K_\psi^{(\xi)-1})_{MP}$ is a (gauge-fixed) gravitino propagator and $\cdots$ includes parts where $Q_M$ in \eqref{eq:L_psi}, \eqref{eq:L_lambda} is involved.
When two D-instantons are widely seperated, the leading order contribution comes from the first term, $(i \Gamma^0\Gamma^M \partial_M )^{-1}$, and we obtain
\begin{align}
  \llangle X^2 \rrangle_{\hbar \rightarrow 0} = \mathcal{C}
\int d^{10}x d^{10}y \, (D_M \bar{\Theta}(x)) P_{cR} (x) \Gamma^R \Gamma^M
\frac{i \Gamma^J(x_J-y_J)}{|x-y|^{10}} \Gamma^N \Gamma^Q P_{cQ} (y)
(D_N \Theta(y)) \,,
\label{eq:X_sq1}
\end{align}
where $\mathcal{C}$ is a numerical constant.
Here, the scalar funciton $P_{cN}(x)$ can be interpreted as a position dependent string coupling, and the derivative couplings on the fermionic degrees of freedom have been exhibited, as we anticipate from the Nambu-Goldstone phenomenon.
The comparison with \eqref{eq:fermion_fluc_quad} tells that $D_N \Theta$ is physically the longitudinal component of $\psi_{\text{(MW)}N}$ in this approximation.

\paragraph{Conclusion and discussion}

In this letter, we have identified local supersymmetry transformation parameteres at the positions of D-instantons with their fermionic coordinates.
We have shown that the two-point function of supercurrents from broken local supersymmetry supplies a tree level (cylinder) contribution to two D-instantons at the boundaries, which exponentiates to give $S_\text{eff}$.
On the other hand, unbroken local supersymmetry does not produce such one and starts instead at one-loop level.
The gravitinos play a few roles, albeit being indirect at this moment, including introducing a multiple of gamma matrices into the effective action.

Extension of our work to higher dimensional D-objects is certainly possible.
For instance, let us consider a Euclidean D1-brane \cite{Green:1996um, Hull:1998vg}.
This object creates a classical profile consisting of a dilaton and a R-R two form, 
and one can carry out the same analysis as is in this letter to obtain an effective action similar to \eqref{eq:X_sq1}.
In these cases, (D-instanton and D1), the fluctuations of D-objects revealed in this letter are not ignorable, and the integrations over $\Theta(X_i)$ are eventually necessary, being treated as ``gas'' \cite{Green:1995my}.
This is in fact a fermionic counterpart of the statement in \cite{Mermin:1966fe, Coleman:1973ci}.


\section*{Acknowledgments}
\label{sec:acknowledgments}

The authors thank A.~Tsuchiya for helpful discussion and Wen-Yu Wen for usuful comments. The work was initiated while two of the authors (H.I. and H.K.) were respectively at National Tsing-Hua University and at National Taiwan University. They acknowledge supports from the universities as well as those by the Ministry of Science and Technology, R.O.C. The work of H.I. is also supported in part by JSPS KAKENHI (23K03393, 23K03394) and H.K. thanks Prof. Shin-Nan Yang and his family for their kind support through the Chin-Yu chair professorship.


\appendix


\end{document}